\documentclass[pre,twocolumn,10pt,aps,showpacs,amsmath,amsfonts,floatfix,letterpaper]{revtex4-1}

\usepackage{times}
\usepackage{txfonts}

\usepackage{amsmath}
\usepackage{amssymb}
\usepackage{hyperref}
\usepackage{graphicx}

\newcommand{\ee}{\mathrm{e}}
\newcommand{\ii}{\mathrm{i}}
\newcommand{\dd}{\mathrm{d}}

\newcommand{\rv}{\boldsymbol{r}}
\newcommand{\qv}{\boldsymbol{q}}
\newcommand{\Phiv}{\boldsymbol{\Phi}}
\newcommand{\phiv}{\boldsymbol{\phi}}
\newcommand{\deltav}{\boldsymbol{\delta}}

\newcommand{\zerov}{\boldsymbol{0}}

\newcommand{\scF}{\mathcal{F}}
\newcommand{\scG}{\mathcal{G}}
\newcommand{\dsZ}{\mathbb{Z}}
\newcommand{\scC}{\mathcal{C}}
\newcommand{\scZ}{\mathcal{Z}}
\newcommand{\bigO}{\mathcal{O}}
\newcommand{\Tm}{\mathbf{T}}
\newcommand{\xhv}{\hat{\boldsymbol{x}}}

\newcommand{\beq}[1]{\begin{equation}\label{#1}}
\newcommand{\eeq}{\end{equation}}
\newcommand{\refeq}[1]{Eq.~(\ref{#1})}

\newcommand{\refeqand}[2]{Eqs.~(\ref{#1}) and (\ref{#2})}
\newcommand{\refcite}[1]{Ref.~\cite{#1}}

\newcommand{\reffig}[1]{Fig.~\ref{#1}}

\newcommand{\refsec}[1]{Section~\ref{#1}}

\newcommand{\refapp}[1]{Appendix~\ref{#1}}

\newcommand{\punc}[1]{\,{\text{#1}}}
\newcommand{\sub}[1]{_{\text{#1}}}

\newcommand{\super}[1]{^{\text{#1}}}

\newcommand{\blp}{\boldsymbol{(}}
\newcommand{\brp}{\boldsymbol{)}}

\DeclareMathOperator{\Div}{div}
\DeclareMathOperator{\Curl}{curl}

\newcommand{\heading}[1]{{\bf \em #1}:}

\usepackage{color}

\begin{document}

\title{Emergence of cooperative dynamics in fully packed classical dimers}

\author{Tom Oakes}
\author{Juan P. Garrahan}
\author{Stephen Powell}
\affiliation{School of Physics and Astronomy, The University of Nottingham, Nottingham, NG7 2RD, United Kingdom}

\begin{abstract}

We study the behavior of classical dimer coverings of the square lattice---a paradigmatic model for systems subject to constraints---evolving under local stochastic dynamics, by means of Monte Carlo simulations and theoretical arguments.  
We observe clear signatures of correlated dynamics in both global and local observables and over a broad range of time scales, indicating a breakdown of the simple continuum description that approximates well the statics.   We show that this collective dynamics can be understood 
in terms of one-dimensional ``strings'' of high mobility, which govern both local and long-wavelength dynamical properties.  We introduce a coarse-grained description of the strings, based on the Edwards--Wilkinson model, which leads to exact results in the limit of low string density and provides a detailed qualitative understanding of the dynamics in all flux sectors.  We discuss the implications of our results for the dynamics of constrained systems more generally.

\end{abstract}

\maketitle

\section{Introduction}
\label{SecIntroduction}

Dimer models are archetypal systems for the study of the effects of strong local constraints \cite{Fisher1961,KenyonReview}. Despite their simplicity, classical dimer models on bipartite lattices exhibit a number of interesting phenomena, such as macroscopic ground-state degeneracy, topological order, and deconfinement of monomers \cite{HenleyReview}. Their static properties are well understood, and are captured by an effective coarse-grained theory, involving a height field \cite{Blote} in two dimensions (2D) or an effective gauge field in higher dimensions \cite{Huse}. In either case, the result is a critical equilibrium phase, with power-law correlations between local degrees of freedom.  This class of systems provides the simplest examples of ``exotic'' thermodynamic behavior purely determined by entropy \cite{HenleyReview}.

Even when the thermodynamic properties of a system have a simple effective description, its dynamics can be more intricate and interesting \cite{Biroli2013}. It is natural to ask whether this is the case for dynamical extensions of the classical dimer model, about which much less is known. 
In particular, Ref.\ \cite{Henley1997}
considered the simplest extension of the coarse-grained description to dynamics, predicting simple relaxational decay of correlations, while Ref.\ \cite{Das2005} considered the dimer model with nonlocal loop dynamics. 
These works should be contrasted with studies of defect-driven dynamics in dimer models \cite{GarrahanPNAS} and of monopole dynamics in spin ice \cite{Jaubert2009,CastelnovoReview}. 

In this work, we consider {\em local} stochastic dynamics in the defect-free square-lattice dimer model, using both simulations and theoretical arguments.  As far as we are aware, what we report here are the first systematic simulation results for the natural dynamics in these systems, i.e., one of locally flipping plaquettes.  We show that the simple continuum picture can fail to describe the true physics even over long time scales and that the phenomenology is, in fact, far richer than the simplicity of the model would suggest. 

The first main contribution of this paper is to demonstrate, using simulations, significant deviations from exponential relaxation in global and local observables over a broad range of time scales. We argue that a simple continuum description fails because the dynamics is {\em facilitated} by local objects---in this case, one-dimensional {\em strings} \cite{Bhattacharjee,Jaubert,Otsuka}---and hence highly heterogeneous. The understanding of the importance of these objects, which has broad implications for the study of cooperative dynamical phenomena, is our second main contribution.

Close-packed dimer models obey a topological constraint \cite{HenleyReview} that amounts to conservation of strings, or, equivalently, of the {\em flux} of an effective magnetic field. Any local rearrangement of dimers conserves flux, and we exploit this by considering dynamics within a fixed flux sector. At large flux, the system is spanned by a low density of strings, whose fluctuations govern the relaxation. We introduce a coarse-grained description of the strings, based on the Edwards--Wilkinson model of fluctuating interfaces \cite{EdwardsWilkinson}, from which we derive exact expressions for dynamical observables, constituting the third main contribution of this work. We confirm these results using simulations at large flux, to which they can be compared with at most one adjustable parameter.

We find that the behavior is qualitatively similar, and consistent with the string picture, for smaller flux, and even in the isotropic limit of vanishing flux. In these latter cases, strings can still be defined and the string picture remains illuminating, even though their density is so high that they cannot be treated as independent. By application of dynamical scaling theory, we furthermore predict a crossover to the Coulomb-phase results of \refcite{Henley1997} at a time scale that diverges on approaching the critical point at saturation flux (zero string density).

\subsection*{Outline}

In \refsec{SecModel}, we introduce the dimer model and the local dynamics that we study in the remainder of the paper. We also briefly review, in \refsec{SecCoarseGrainedHeight}, the coarse-grained theory introduced by Henley \cite{Henley1997} to describe dynamics of the height field, which predicts exponential decay of correlations. The majority of our original results are presented in \refsec{SecStringDynamics}, where we use an effective theory of string dynamics based on the Edwards--Wilkinson equation to derive results for correlations and for the persistence, a local probe of dynamics. We conclude in \refsec{SecConclusions} with a brief discussion of the broader significance of our results for dynamics in strongly constrained systems. Some technical details and additional simulation results are presented in appendices.

\section{Model}
\label{SecModel}

\begin{figure*}
\begin{center}
\includegraphics[width=\textwidth]{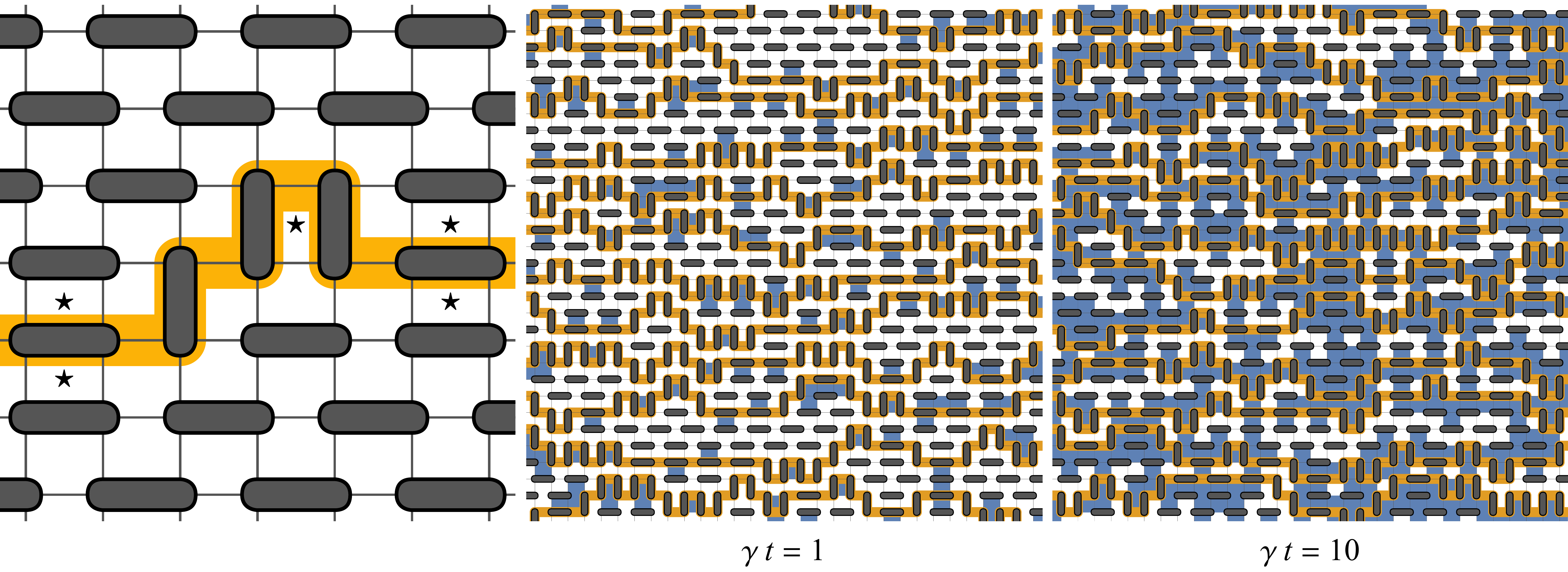}
\end{center}
\caption{Left: A dimer configuration with a single string, relative to a fully staggered configuration with maximal flux \(\Phi_x = \frac{1}{2}L^2\) along the horizontal direction. A single string spanning the system once reduces the flux \(\Phi_x\) by \(-L\), the smallest possible amount, irrespective of its path. Flippable plaquettes, which appear when the dimers are shifted, are marked with stars (\(\star\)); the fully staggered configuration has none. Center and right: Configurations with \(\phi_x = \frac{1}{4}\) evolved for time \(t\).  Persistent plaquettes are white, while those that have flipped are blue; strings are yellow. The dynamics is spatially heterogeneous: Even at relatively long times, extended regions are unvisited by strings and hence persistent.}
\label{FigStrings}
\end{figure*}
We study a dimer model on an \(L\times L\) square lattice with periodic boundaries. The occupation variable \(d_\mu(\rv)\) gives the number, \(0\) or \(1\), of dimers on the link joining sites \(\rv\) and \(\rv + \deltav_\mu\), where \(\mu\in\{x,y\}\) and \(\deltav_\mu\) is a lattice vector. A configuration is allowed only if every site is occupied by a single dimer. We refer to a plaquette as {\em flippable} when it contains two parallel dimers; the {\em flippability} \(f\) of a configuration is defined as the proportion of plaquettes that are flippable,
\begin{equation}
f = \frac{1}{L^2}\sum_{\rv} \sum_{\mu\nu} d_\mu(\rv) d_\mu(\rv + \deltav_\nu)\punc{.}
\end{equation}

We define the effective ``magnetic field'' \(B_\mu(\rv) = \varepsilon_{\rv}[d_\mu(\rv) - \frac{1}{4}]\), where \(\varepsilon_{\rv} = \pm 1\) on the two sublattices \cite{HenleyReview}. The constraint on dimer configurations then becomes Gauss' law, \(\Div_{\rv} B = 0\), where
\begin{equation}
\Div_{\rv} B = \sum_\mu \left[B_\mu(\rv) - B_\mu(\rv - \deltav_\mu)\right]
\end{equation}
is the lattice divergence. The flux \(\Phiv\) corresponding to \(B\) can be defined by \(\Phi_\mu = \sum_{\rv} B_{\mu}(\rv) = \sum_{\rv} \varepsilon_{\rv} d_\mu(\rv)\). Because \(\varepsilon_{\rv + \deltav_\nu} = -\varepsilon_{\rv}\), a pair of neighboring parallel dimers, of either orientation, gives zero net contribution to \(\Phiv\), and so plaquette-flip dynamics conserves the flux.

\subsection{Staggered configurations and strings}
\label{SecStaggeredConfigurations}

The flux is maximized by a staggered dimer configuration. For example, if \(d_y(\rv) = 0\) for all \(\rv\), \(d_x(\rv) = 1\) for \(\varepsilon_{\rv} = +1\), and \(d_x(\rv) = 0\) otherwise, then \(\Phiv = +\Phi\sub{max} \xhv\), where \(\Phi\sub{max} = \frac{1}{2}L^2\). The other three staggered configurations, related by symmetry, have flux of the same magnitude, \(\lvert\Phiv\rvert = \Phi\sub{max}\), along other lattice directions. We define the flux relative to its maximum by \(\phiv = {\Phiv}/{\Phi\sub{max}}\) and, for flux along the \(x\) direction, the deviation from maximum \(\theta = 1 - \phi_x\).

To reduce the flux from maximum, one can shift a row of dimers spanning the system, which changes \(\Phi_x\) by \(-L\). We refer to such a set of shifted dimers as a string \cite{Jaubert,SpinIceCQ}. After the shift, plaquettes along its length become flippable; flipping these deforms the string but conserves the flux. A possible path for a single string and the resulting configuration are shown in \reffig{FigStrings}. With \(N\sub{s}\) strings introduced into a staggered configuration, \(\theta = {2N\sub{s}}/{L}\); the linear density is therefore \(\frac{1}{2}\theta\).

\subsection{Height mapping}

The constraint \(\Div_{\rv} B = 0\) can be resolved by defining the {\em height} \(z\) on each plaquette  \cite{Henley1997}, in terms of which \(B_\mu(\rv) = -\frac{1}{4}\Curl_{(\rv,\mu)} z\), where the curl is the difference between the plaquettes on each side of a link. Global shifts of \(z\) do not affect \(B\), corresponding to the gauge redundancy in 3D \cite{HenleyReview}. With an appropriate gauge choice, flipping a plaquette modifies \(z\) only on that plaquette \cite{Henley1997}. If \(B\) has periodic boundary conditions, \(z(\rv + L \deltav_\mu) = z(\rv) + 4L^{-1}\sum_\nu\epsilon_{\mu\nu}\Phi_\nu\), where \(\epsilon\) is the Levi-Civita tensor. The spatial average of the derivative of the height (the {\em tilt}) is therefore intensive and proportional to \(\phiv\). We define \(\zeta(\rv) = z(\rv) - 2\sum_{\mu\nu}\epsilon_{\mu\nu} r_\mu \phi_\nu\), with periodic boundary conditions.

\subsection{Dynamics}

The most natural dynamics for the dimer system is one where individual plaquettes flip randomly. This dynamics is efficiently implemented numerically via continuous-time Monte Carlo (MC) \cite{Newman}, in which, when flippable, plaquettes flip according to a Poisson process with rate constant \(\gamma\). Dynamics at equilibrium within a sector of fixed flux \(\phiv\) can be studied by starting from a fixed configuration and equilibrating using plaquette-flip dynamics. We will denote by \(\langle\cdots\rangle\) an average both over the equilibrium ensemble (where all allowed states with flux \(\phiv\) have equal weight) and, where applicable, over subsequent trajectories.

The dynamical correlation function of the height is defined by \(G_\zeta(\qv, t) = \langle \tilde{\zeta}(\qv,t) \tilde{\zeta}(-\qv,0)\rangle\) where \(\tilde{\zeta}\) is the Fourier transform of \(\zeta\). We also consider the persistence \(p(t)\), the proportion of plaquettes that have not flipped at any point up to time \(t\), which provides a local probe of the evolution.

\subsection{Continuous height-field theory}
\label{SecCoarseGrainedHeight}

The static properties of the dimer model can be described by a continuum theory in terms of the coarse-grained height \(h(\rv)\) resulting from averaging \(\zeta(\rv)\) over short length scales \cite{HenleyReview}. Apart from terms irrelevant at long distances, the effective dimensionless free energy is
\begin{equation}
\label{EqContinuumFh}
\scF = \frac{1}{2}\int \dd^2 \rv [ K_y(\nabla_x h)^2 + K_x(\nabla_y h)^2 ]
\end{equation}
for \(\phiv\) along \(\xhv\), implying correlations \(\langle \tilde{h}(\qv) \tilde{h}(-\qv) \rangle = [\omega(\qv)]^{-1}\) for the Fourier transform \(\tilde{h}\), where \(\omega(\qv) = K_y q_x^2 + K_x q_y^2\).

The simplest extension of the continuum description to dynamical properties is the Langevin equation \cite{Henley1997}
\begin{equation}
\label{EqLangevin}
\frac{\partial}{\partial t} h(\rv,t) = -\Gamma \frac{\delta \scF}{\delta h(\rv,t)} + \eta_h(\rv,t)\punc{,}
\end{equation}
where the noise has correlations \(\langle \eta_h(\rv,t)\eta_h(\rv',t')\rangle = 2\Gamma \delta(\rv - \rv')\delta(t-t')\). The resulting two-time correlations are \cite{Henley1997}
\beq{EqDynCorrnsh}
G_h(\qv,t) \equiv \langle \tilde{h}(\qv, t) \tilde{h}(-\qv,0)\rangle = \frac{\ee^{-\Gamma \omega(\qv) t}}{\omega(\qv)}
\punc{,}
\eeq
implying exponential decay at long time scales. We show below that this prediction can break down, even when the height-field approach is accurate for the statics, due to cooperative effects that dominate the dynamics.

\section{String dynamics}
\label{SecStringDynamics}

The collective character of the dynamics can be uncovered by considering the behavior near maximal flux (i.e., for small \(\theta\)),  where most of the system is unflippable.  This regime can be understood in terms of a low density of well-separated strings. We first consider the dynamics of a single string, using a continuum description based on the Edwards--Wilkinson equation, before turning to the consequences for the two classes of observables, correlation functions and the persistence.

For a single string traversing the system horizontally, let \(y(x,t)\) be the vertical position at horizontal position \(x\) and time \(t\). By using a transfer matrix to enumerate all possible string configurations (see \refcite{Otsuka} and \refapp{AppTransferMatrix}), we find the following two exact results regarding the equilibrium distribution of a single string: (i) The mean number of flippable plaquettes is given by
\beq{EqNf}
N_f = \left(2 - \sqrt{2}\right)L + \bigO(\ln L)\punc{.}
\eeq
(ii) For \(1 \ll \lvert x - x'\rvert \ll L\), \(y(x,t) - y(x',t)\) is normally distributed with zero mean and variance \( \lvert x - x'\rvert / {\sqrt{2}} \).

\subsection{Edwards--Wilkinson equation}

At large length and time scales, we expect \(y(x,t)\) to obey the Edwards--Wilkinson equation \cite{EdwardsWilkinson,Bray},
\beq{EqEW}
\frac{\partial}{\partial t} y(x,t) = \frac{1}{2}\Lambda \frac{\partial^2}{\partial x^2} y(x,t) + \eta(x,t)\punc{,}
\eeq
where \(\langle \eta(x,t) \eta(x',t') \rangle_0 = D\Lambda\delta(x-x')\delta(t-t')\), and \(\Lambda\) and \(D\) parameterize, respectively, the stiffness of the string and the strength of the noise. The average \(\langle\cdots\rangle_0\) is taken over trajectories starting from a given initial configuration \(y(x,0)\).

Accounting for the periodicity in the \(x\) direction (but not in \(y\)), the Green function for \refeq{EqEW} is
\begin{equation}
\Delta(x,t) = \frac{1}{L}\sum_{k} \ee^{\ii k x} \ee^{-\frac{1}{2}\Lambda t k^2}\punc{,}
\end{equation}
where \(k L/2 \pi  \in \dsZ\). To calculate the two-time correlation function in the equilibrium ensemble, we take both times to infinity with their difference finite,
\begin{align}
V_L(x, t) &\equiv \left\langle [y(x,t)-y(0,0)]^2\right\rangle\\
&= \lim_{t_0 \rightarrow \infty} \left\langle [y(x,t_0+t)-y(0,t_0)]^2\right\rangle_0
\nonumber \\
&=\frac{D \Lambda t}{L} + \frac{2D}{L}\sum_{{k\neq 0}}\frac{1-\ee^{\ii k x}\ee^{-\frac{1}{2}\Lambda t k^2}}{k^2}\punc{.}
\label{EqDefineV}
\end{align}

The typical width of a string in equilibrium can be characterized by the mean-square displacement between the points \(x\) and \(0\) at equal time, which is given, for \(0\le x\le L\), by \(V_L(x,0) = D x (1 - {x}/{L})\). The time scales for dynamics can similarly be understood through \(V_L(0,t)\), which is shown in \refapp{SecStringCoarseGrained} to obey
\beq{EqStringDisplacementVariance}
\left\langle [y(0,t)-y(0,0)]^2\right\rangle \approx \begin{cases}
D\sqrt{\frac{2\Lambda t}{\pi}}&\text{for \(\Lambda t \ll L^2\)}\\
\frac{D}{L}\Lambda t&\text{for \(\Lambda t \gg L^2\).}
\end{cases}
\eeq
The short-time result, \(\Lambda t \ll L^2\), gives the dynamical scaling relation between the characteristic length in the \(y\) direction and time through the ``growth exponent'' \(\beta\) \cite{Bray},
\beq{EqScaling}
l_{y} \sim t^{\beta}\punc{,}
\qquad
\beta = \frac{1}{4}\punc{.}
\eeq
At long times, \(\Lambda t \gg L^2\), the whole string can be treated as a random walker, with effective diffusion constant \({D \Lambda}/{L}\).

These results, along with those from the string microscopics, fix the values of \(D\) and \(\Lambda\). Comparison of \(V_L(x,0)\) with the exact result for the equal-time displacement variance gives \(D = 1/ \sqrt{2} \). The mean rate of plaquette flips in equilibrium is \(\gamma N_f\).  Since each flip changes the mean vertical position by \(\pm L^{-1}\), the variance of the total shift is \(\gamma t N_f/L^2\) in the long-time limit. Comparison of \refeqand{EqNf}{EqStringDisplacementVariance} therefore gives \(\Lambda = 2(\!\sqrt{2} - 1)\gamma\).

\subsection{Height correlation function}

To calculate height correlations based on the coarse-grained string description, we write
\begin{equation}
\nabla_y h(\rv,t) \propto \int_0^L \dd x\,\delta^2\blp\rv - \{x,y(x,t)\}\brp - \frac{1}{L}\punc{,}
\end{equation}
which treats the string as a step in the height plus a uniform gradient to preserve the boundary conditions. The height correlation function for \(\qv \neq \zerov\) can then be written, for a single string, as
\(G_h(\qv,t) \propto (L q_y^2)^{-1} C\sub{s}(\qv,t)\),
where
\begin{equation}
C\sub{s}(\qv,t) = \frac{1}{L}\int_0^L \dd x \int_0^L \dd x'\,\ee^{-\ii q_x (x-x')}\left\langle\ee^{-\ii q_y [y(x,t)-y(x',0)]} \right\rangle\punc{.}
\nonumber
\end{equation}
For small-enough density \(\frac{1}{2}\theta\), string contributions add incoherently, resulting in \(G_h(\qv,t) \propto \theta q_y^{-2} C\sub{s}(\qv,t)\). Since \(y(x,t)-y(x',0)\) is Gaussian distributed with zero mean, we get
\begin{equation}
\label{EqCsFull}
C\sub{s}(\qv,t) = \int_{-\frac{L}{2}}^{\frac{L}{2}} \dd x \,\ee^{-\ii q_x x}\ee^{-\frac{1}{2}q_y^2 V_L(x,t)}\punc{,}
\end{equation}
where the periodicity of \(V_L(x,t)\) under \(x \rightarrow x \pm L\) has been used to shift the limits of integration.

\begin{figure}
\begin{center}
\includegraphics[width=\columnwidth]{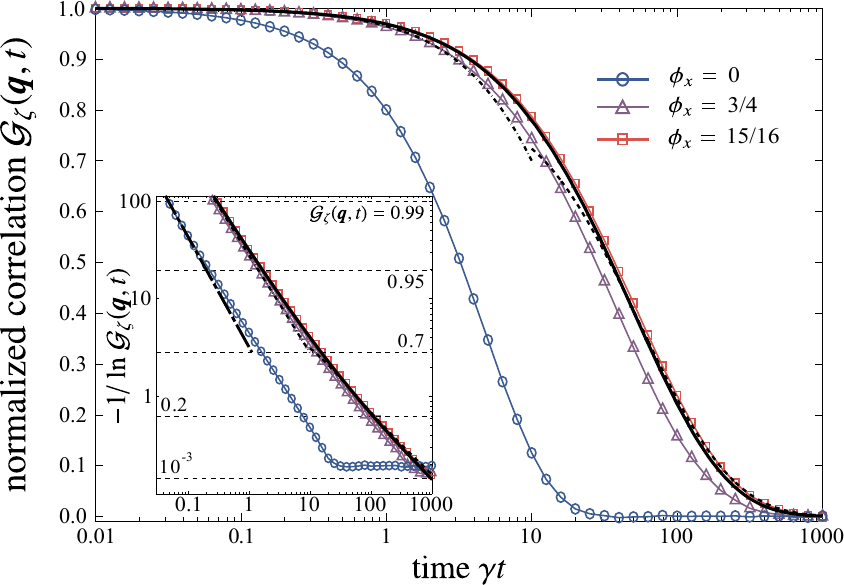}
\end{center}
\caption{Normalized height correlations \(\scG_\zeta(\qv,t) = G_\zeta(\qv,t)/G_\zeta(\qv,0)\) at wavevector \(\qv = \{\frac{\pi}{16},\frac{\pi}{4}\}\) and flux \(\phiv = \{\phi_x,0\}\). Symbols show MC results (error bars are smaller than symbols), while the thick (black) line is the theoretical prediction for the large-flux limit (i.e., close to the maximum \(\phi_x = 1\)). The short- and long-time limits, \refeq{EqDynCorrnsString}, are shown with dot-dashed and dashed lines, respectively. The system size is \(L = 256\); besides restricting \(\phi_x\) to discrete values, finite-size effects are minimal. Inset: Same data with double-logarithmic vertical scale. On this plot, a stretched exponential \(\ee^{-(t/\tau)^\beta}\) appears as a straight line with slope \(-\beta\). A dot--dashed straight line with slope \(1\) is shown for comparison; the data deviate from this slope, indicating stretching, even for zero flux.}
\label{FigCorrelations}
\end{figure}

Asymptotic expressions for the correlations can be found in various limits. For the static correlations, \(G_h(\qv,0) \propto \theta/\omega\sub{s}(\qv)\), where \(\omega\sub{s}(\qv) = q_x^2 + \frac{1}{8}q_y^4\) in the thermodynamic limit. For \(t \ll L^2\), we find time dependence
\begin{equation}
\label{EqDynCorrnsString}
\frac{G_h(\qv,t)}{G_h(\qv,0)} \propto \begin{cases}
\exp \left[-\frac{1}{2}\Lambda t \, \omega\sub{s}(\qv)\right]&\text{for \(\Lambda t \ll q_y^{-1/\beta}\)}\\
t^\beta
\exp\left[-\sqrt{\Lambda t \, \tilde\omega\sub{s}(\qv)}\right]&\text{for \(q_y^{-1/\beta} \ll \Lambda t\),}
\end{cases}
\end{equation}
where \(\tilde\omega\sub{s}(\qv) = (4\pi)^{-1}q_y^4 \exp[2(\operatorname{erfi}^{-1} \sqrt{8}q_x/q_y^2)^2]\); the proportionality constants are calculated exactly in \refapp{AppDynamicalCorrelations}. A crossover from simple to stretched exponential therefore occurs at \(\Lambda t \sim q_y^{-1/\beta}\), the time scale corresponding, according to \refeq{EqScaling}, to the wavelength \(\sim q_y^{-1}\). The stretching is the result of contributions from the continuum of modes of the string. These expressions, along with the full result found by numerical integration of \refeq{EqCsFull}, are compared with simulations in \reffig{FigCorrelations}. We find close agreement, with no adjustable parameters, at large flux and small wavevector, and qualitative agreement at smaller \(\phi_x\) and larger \(\qv\); see \refapp{AppNumerics}.

The independent-string approximation should be valid for \(\Lambda t \ll \theta^{-1/\beta}\), the time corresponding to a \(y\) displacement equal to the mean string separation. The stretched-exponential form therefore applies up to a time that diverges at \(\theta = 0\). For larger \(t\), string interactions are important for the dynamics, and \refeq{EqDynCorrnsString} is no longer valid. When \(\Lambda t \gg \theta^{-1/\beta}\) many strings contribute, their discreteness becomes unimportant, and so we expect a crossover to the Coulomb-phase behavior of \refeq{EqDynCorrnsh}. These crossovers can be understood via dynamical scaling theory, based on the critical point at \(\theta = 0\) \cite{Jaubert,SpinIceCQ}, whose critical theory is that of hard-core bosons---or, equivalently, free fermions---in 1D. All critical exponents are therefore rational and follow from dimensional analysis.

\subsection{Persistence}

The persistence \(p(t)\) can similarly be understood in terms of the behavior of strings. At very short times, \(p(t) = \ee^{-\langle f \rangle \gamma t}\): plaquettes which are flippable at \(t = 0\) will flip independently with rate $\gamma$.  For times \( \gamma t \gg 1 \), each point \(x\) on a string performs a subdiffusive random walk, according to \refeq{EqStringDisplacementVariance}. As only plaquettes adjacent to strings are flippable, the mean persistence is equal to the probability that a plaquette is not reached by any string up to time \(t\), and is therefore given by the survival probability for a stationary target in the presence of a density \(\frac{1}{2}\theta\) of subdiffusive traps. Using the results of \refcite{FrankeMajumdar} relating the dynamic exponent to the persistence, we get
\begin{equation}
\label{EqPersistenceLargeFlux}
\langle p(t)\rangle \propto \exp \left[-\frac{\theta}{\Gamma(5/4)}(\Lambda t)^\beta\right]\;\text{for \(\gamma^{-1} \ll t \ll \Lambda^{-1}\theta^{-1/\beta}\).}
\end{equation}
This form ceases to apply for \(\Lambda t \sim \theta^{-1/\beta}\), when string interactions become important, or for \(\Lambda t \sim L^2\), beyond which the long-time behavior in \refeq{EqStringDisplacementVariance} applies and \(\langle p(t)\rangle \propto \ee^{-c t^{1/2}}\).

\begin{figure}
\begin{center}
\includegraphics[width=\columnwidth]{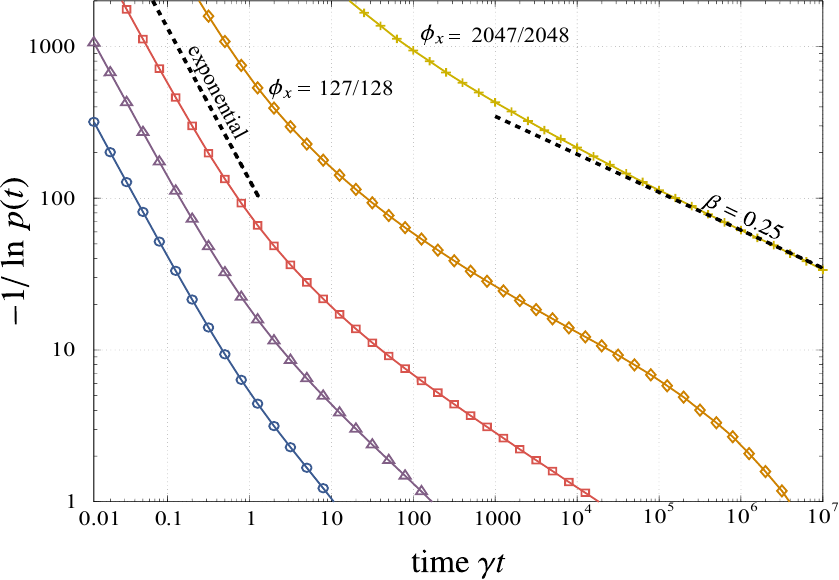}
\end{center}
\caption{MC results for persistence \(p(t)\) at flux \(\phi_x\) labeled as in \reffig{FigCorrelations}, except where indicated. (Error bars are smaller than symbols.) System size \(L=4096\) is required to reach flux \(\phi_x = 2047/2048\) (see \refsec{SecStaggeredConfigurations}); for the other flux values, \(L = 256\). As expected based on the string picture, an initial exponential decay (straight line with slope \(1\)) is followed by a stretched exponential (\(\text{slope} < 1\)), lasting until \(\Lambda t \sim (1-\phi_x)^{-1/\beta}\). The dashed line at the right, with slope \(1/4\), is the function \(p(t) \propto \ee^{-c t^{1/4}}\), in agreement with \refeq{EqPersistenceLargeFlux} for the limit \(\phi_x \rightarrow 1\).}
\label{FigPersistence}
\end{figure}

Our simulation results, \reffig{FigPersistence}, are in qualitative agreement with these arguments, showing an initial exponential followed by a stretched exponential, for all fluxes. The stretching exponent decreases continuously with \(\phi_x\), approaching \(\beta = 1/4\), in agreement with \refeq{EqPersistenceLargeFlux}, as \(\phi_x \rightarrow 1\). At longer times, a faster decay, consistent with a single exponential, is observed.

\section{Conclusions}
\label{SecConclusions}

We have shown that the close-packed square-lattice dimer model, subject to local, plaquette-flip dynamics, displays emergent collective relaxation that is not anticipated by simple extensions of its static properties.  Approximations to the dynamics based on free-energy gradients plus noise, such as \refeq{EqLangevin}, fail to capture the intrinsic heterogeneity: Due to the constrained nature of the system, motion is only allowed in the vicinity of strings, and relaxation is dominated by spatial fluctuations.  In a sense, the noise that triggers rearrangements is not uniform in space and time; rather, its strength depends sensitively on the local configuration.  Strings \emph{facilitate} local rearrangements, dynamics is heterogeneous and collective (see \reffig{FigStrings}), and relaxation functions are nonexponential.

This situation is reminiscent of glass-forming systems \cite{Garrahan}: In a slowly relaxing material such as a glass former, ``facilitation'' indicates the fact that local relaxation can occur only near an already locally relaxing region \cite{Keys2011}. Similarly, in the dimer model, plaquette moves are only possible in the vicinity of a string. This is the reason that the Langevin dynamics of \refeq{EqLangevin} is not accurate for relaxation in regions where string density is low.  The additive noise assumed in that approximation would allow rearrangements to occur anywhere in space. But if dynamics is facilitated, the noise that drives local rearrangements is not uniformly distributed, but rather concentrated near already mobile regions. A Langevin description, along the lines of \refeq{EqLangevin}, would therefore require a form of noise that is multiplicative and whose magnitude is strongly dependent on the local flippability.

Instead we have developed a string description of the dynamics, which directly incorporates the local nature of the relaxation. At low string density (high flux), we are able to make exact theoretical predictions for the correlations and persistence, which are confirmed by our simulations. We in fact find that much of the qualitative behavior is unchanged at smaller flux, where interactions between strings are certainly important. This indicates that the usefulness of the string picture, as well as the concept of facilitated and heterogeneous dynamics, extends well beyond the regime of high flux.

Besides their fundamental importance, our results are likely to be of relevance to spin ice, where closely analogous string excitations have been evidenced directly using neutron scattering \cite{Morris}, and where correlations with stretched-exponential decay have been noted \cite{Revell}. Dynamical results for classical dimers are also relevant to the corresponding quantum dimer model at its Rokhsar--Kivelson point \cite{Henley2004,Laeuchli}.

\acknowledgements

The simulations used resources provided by the University of Nottingham High-Performance Computing Service. This work was supported by EPSRC Grant No.\ EP/K01773X/1 (JPG) and EPSRC Grant No.\ EP/M019691/1 (SP).

\heading{Data availability} Research data are available from the Nottingham Research Data Management Repository at \url{http://dx.doi.org/10.17639/nott.40}.

\appendix

\section{Transfer-matrix calculation of string configurations}
\label{AppTransferMatrix}

\begin{figure}[b]
\begin{center}
\includegraphics[width=0.9\columnwidth]{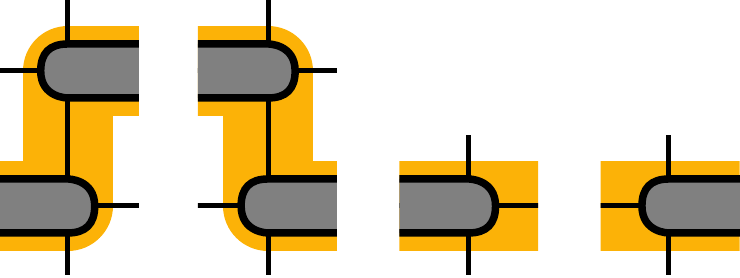}
\end{center}
\caption{Four types of segment that can be combined to form a string. The first two are steps at which the vertical position of the string changes by \(\pm 1\), while the last two are the two segments of a horizontal step. (This step is split so that all four segments involve the same horizontal displacement.)}
\label{FigStringSteps}
\end{figure}

A string can be divided into four types of segment, illustrated in \reffig{FigStringSteps}; the ensemble \(\scC\super{s}_L\) of configurations for a single string is given by the set of ways in which these segments can be combined to produce a closed path of length \(L\).
One can write a transfer matrix
\beq{EqTransferMatrix}
\Tm(k,\mu) = \begin{pmatrix}
\ee^{-\ii k}&\ee^{-\ii k}\ee^{-\mu}&0&\ee^{-\ii k}\\
\ee^{\ii k}\ee^{-\mu}&\ee^{\ii k}&0&\ee^{\ii k}\\
1&1&0&1\\
0&0&\ee^{-2\mu}&0
\end{pmatrix}\punc{,}
\eeq
such that \(T_{\beta\alpha}(k,\mu)\) is nonzero only when a segment of type \(\alpha\) can be followed by one of type \(\beta\) (labeled according to the order in \reffig{FigStringSteps}). Each successive pair of segments is weighted by \(\ee^{-\mu}\) for every flippable plaquette it produces and by \(\ee^{-\ii k}\) for every step in the positive \(y\) direction. We define the partition function \(\scZ\super{o}_L\) as the weighted sum over all \emph{open} paths of length \(L\) with \emph{any} net vertical displacement \(y\),
\beq{EqPartitionFunckmu}
\scZ\super{o}_L(k,\mu) = \sum_{\scC\super{o}_L} \ee^{-\ii k y} \ee^{-\mu N_{f}}\punc{,}
\eeq
where \(\scC\super{o}_L\) denotes the ensemble of such paths and \(N_f\) is the number of flippable plaquettes in the resulting dimer configuration. Summing over all sequences of path segments and (vertical) starting positions gives \footnote{Summing over all starting positions for paths results in a factor of \(L\), but this effectively includes both (horizontal) fully staggered configurations; the factor of \(\frac{1}{2}\) corrects the double counting.}
\begin{align}
\scZ\super{o}_L(k,\mu) &= \frac{L}{2}\operatorname{Tr}[\Tm(k,\mu)]^L\\
&= \frac{L}{2}\sum_{\lambda\in\sigma_{\Tm(k,\mu)}}\lambda^L
\punc{,}
\end{align}
where the sum is over the set \(\sigma_{\Tm(k,\mu)}\) of eigenvalues \(\lambda\) of \(\Tm(k,\mu)\).

The allowed paths for a single string are those that return to their starting point after winding once around the system, and hence have net displacement \(y = 0\). The partition function for such paths is
\begin{align}
\label{EqPartitionFuncmu}
\scZ\super{s}_L(\mu) &= \sum_{\scC\super{s}_L}\ee^{-\mu N_{f}} \\
&= \sum_{\scC\super{o}_L} \delta_{y,0} \ee^{-\mu N_{f}}
= \int_{-\pi}^{\pi}\frac{\dd k}{2\pi} \sum_{\scC\super{o}_L} \ee^{-\ii k y} \ee^{-\mu N_{f}}\\
&= \frac{L}{2}\int_{-\pi}^{\pi}\frac{\dd k}{2\pi} \sum_{{\lambda\in\sigma_{\Tm(k,\mu)}}}\lambda^L\punc{.}
\end{align}
For large \(L\), the saddle-point approximation gives
\begin{equation}
\ln \scZ\super{s}_L(\mu) = L \ln \lvert\lambda\sub{max}(\mu)\rvert + \bigO(\ln L)\punc{,}
\end{equation}
where
\begin{align}
\lvert\lambda\sub{max}(\mu)\rvert &= \max_k \lvert\lambda\sub{max}(k,\mu)\rvert\\
&= \max_k \max \{ \lvert\lambda\rvert : \lambda \in \sigma_{\Tm(k,\mu)}\}
\end{align}
is the largest eigenvalue (by magnitude) of \(\Tm(k,\mu)\) for any \(k\). The maxima are \(\lambda\sub{max}(\mu) = 1 + \sqrt{2} - \sqrt{2}\mu + \bigO(\mu^2)\), occurring at the points \(k = 0\) and \(\pm \pi\), and hence
\begin{equation}
\frac{1}{L}\ln \scZ\super{s}_L(\mu) \simeq \ln (1+\sqrt{2}) - (2-\sqrt{2})\mu\punc{.}
\end{equation}

Setting \(\mu = 0\) gives the entropy of a single string,
\begin{align}
S\super{s}_L &= \ln \scZ\super{s}_L(0)\\
&=L \ln (1+\sqrt{2}) + \bigO(\ln L)\punc{.}
\end{align}
The mean number of flippable plaquettes in the presence of a single string is given by
\begin{align}
N\super{s}_f &= \!\left.-\frac{\dd}{\dd \mu} \ln \scZ\super{s}_L(\mu)\right\rvert_{\mu = 0}\\
&= (2 - \sqrt{2})L + \bigO(\ln L)\punc{.}
\label{EqExactMeanFlippable}
\end{align}
At flux \(\phiv = \{1 - \theta,0\}\), the number of strings is \(N\sub{s} = \frac{1}{2}L\theta\). If the strings can be treated as approximately independent, as expected for sufficiently small \(\theta\), then the number of flippable plaquettes is simply \(N\sub{s}N_f\super{s}\), and so the mean flippability in equilibrium is
\begin{equation}
\langle f \rangle = \left(1 - \frac{1}{\sqrt{2}}\right)\theta + \bigO\left(\frac{\ln L}{L}\right)\punc{.}
\label{EqFlippability}
\end{equation}

\begin{figure}
\begin{center}
\includegraphics[width=\columnwidth]{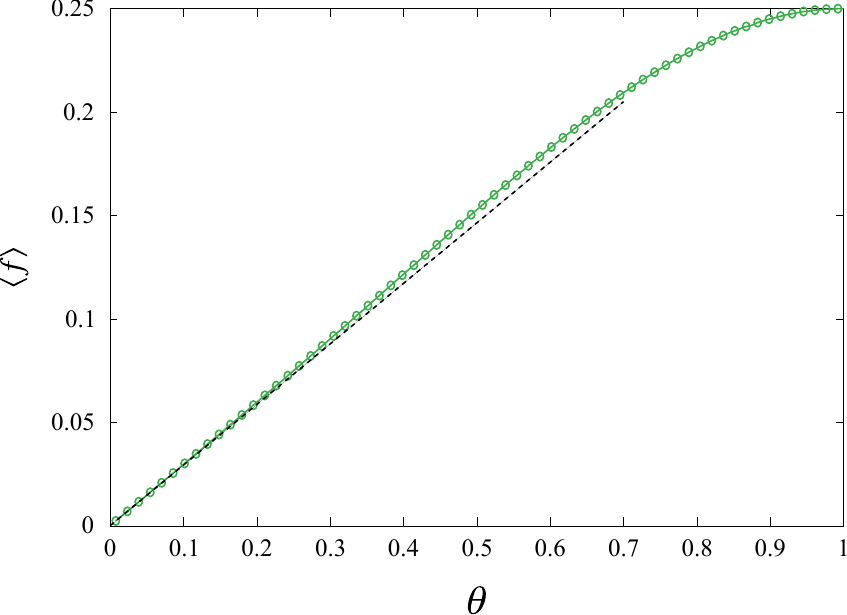}
\end{center}
\caption{Mean flippability in equilibrium \( \langle f \rangle \) as a function of deviation from maximum flux, \(\theta = 1 - \phi_x\). The solid (green) line shows Monte Carlo results for system size \(L=256\) (error bars are smaller than symbols), while the dashed (black) line shows the analytical prediction from \refeq{EqFlippability}.}
\label{FigFlippability}
\end{figure}
Numerical results, shown in \reffig{FigFlippability}, confirm \refeq{EqFlippability} in the limit of small \(\theta\) and are in approximate agreement even for fairly large \(\theta\), suggesting that the independent-string picture is reasonable. (The logarithmic corrections modify the coefficient of \(\theta\), and are of the order of a few percent for \(L = 256\).)

The width distribution of a single string can be determined by a similar approach: The probability distribution for the net vertical displacement \(Y\) of a section of string with horizontal extent \(X\) is \footnote{The ensemble \(\scC\super{o}_X\) has the additional constraint that the last and first segments of the open path are compatible, which allows \(\scZ\super{o}_X\) to be expressed as a trace. This has no effect on the large-\(X\) behavior of \(P\super{s}(X,Y)\).}
\begin{align}
P\super{s}(X,Y) &= \frac{\sum_{\scC\super{o}_X} \delta_{y,Y}}{\scZ\super{o}_X(0,0)}\\
&= \int_{-\pi}^{\pi}\frac{\dd k}{2\pi} \,\ee^{\ii k Y}\frac{\sum_{{\lambda\in\sigma_{\Tm(k,0)}}}\lambda^X}{\sum_{\lambda\in\sigma_{\Tm(0,0)}}\lambda^X}\punc{.}
\end{align}
For large \(X\), the ratio of sums can be found by expanding \(\lambda\sub{max}(k,0)\) in a Taylor expansion around its maxima, giving a pair of Gaussians of variance \(\frac{\sqrt{2}}{X}\) centred at \(k = 0\) and \(\pm\pi\). Taking the Fourier transform, one finds that \(P\super{s}(X,Y)\) is given by a normal distribution of variance \(\frac{1}{\sqrt{2}}X\) when \(X\) and \(Y\) have the same parity, and vanishes otherwise (as required by the structure of a string).

If, for a single string traversing the system in the horizontal direction, we denote by \(y(x)\) the vertical position at horizontal position \(x\), this result can be restated as follows: At length scales much larger than the lattice scale but smaller than the system size, \(1 \ll \lvert x - x'\rvert \ll L\), the vertical displacement \(y(x) - y(x')\) is normally distributed with zero mean and
\beq{EqExactStringWidth}
\left\langle[y(x) - y(x')]^2\right\rangle = \frac{1}{\sqrt{2}}\lvert x - x'\rvert\punc{.}
\eeq
This result is confirmed by the small-displacement limit in \reffig{FigEquilibriumStringWidth}.

\section{Coarse-grained string picture}
\label{SecStringCoarseGrained}

The Edwards--Wilkinson equation, \refeq{EqEW}, has general solution (for \(t \ge 0\))
\begin{multline}
y(x,t) = \int_0^L\! \dd x'\,\Delta(x-x',t)y(x',0)\\
 + \int_0^L\! \dd x'\int_0^\infty \dd t'\, \Delta(x-x',t-t')\eta(x',t')
\end{multline}
(accounting for the periodicity in the \(x\) direction, but not in the \(y\) direction), where
\beq{EqDeltaPropagator}
\Delta(x,t) =
\begin{cases}
{L^{-1}}\sum_{{k \in \frac{2\pi}{L}\dsZ}} \ee^{\ii k x} \ee^{-\frac{1}{2}\Lambda t k^2} & \text{for \(t \ge 0\)}\\
0 & \text{for \(t < 0\)}
\end{cases}
\eeq
is the retarded propagator.

\begin{widetext}
The two-time correlation function, within an ensemble of trajectories with fixed initial configuration, is therefore given by
\begin{multline}
\label{EqCorrelatorWithFixedInitialConfig}
\left\langle [y(x,t)-y(x',t')]^2\right\rangle_0 = \left\{\int_0^L\! \dd x''\,\left[\Delta(x-x'',t)-\Delta(x'-x'',t')\right]y(x'',0)\right\}^2\\
{}+D \Lambda \int_0^L\! \dd x''\int_0^\infty \dd t''\, \left[\Delta(x-x'',t-t'') - \Delta(x'-x'',t'-t'')\right]^2\punc{.}
\end{multline}
To calculate the equivalent correlation function in an equilibrium ensemble, \(\langle\cdots\rangle\), we take both times to infinity while keeping their difference finite:
\begin{equation}
\left\langle [y(x,t)-y(x',t')]^2\right\rangle = \lim_{t_0 \rightarrow \infty} \left\langle [y(x,t_0+t)-y(x',t_0+t')]^2\right\rangle_0\punc{.}
\end{equation}
In this limit, \(\Delta(x,t_0+t) = L^{-1}\), and so the first term in \refeq{EqCorrelatorWithFixedInitialConfig}, which depends on the initial configuration, vanishes. The integrals in the second term can be performed to give
\begin{equation}
\left\langle [y(x,t)-y(x',t')]^2\right\rangle = V_L(\lvert x-x' \rvert,\lvert t-t'\rvert)\punc{,}
\end{equation}
where \(V_L\) is given in \refeq{EqDefineV}.
\end{widetext}

\begin{figure}
\begin{center}
\includegraphics[width=\columnwidth]{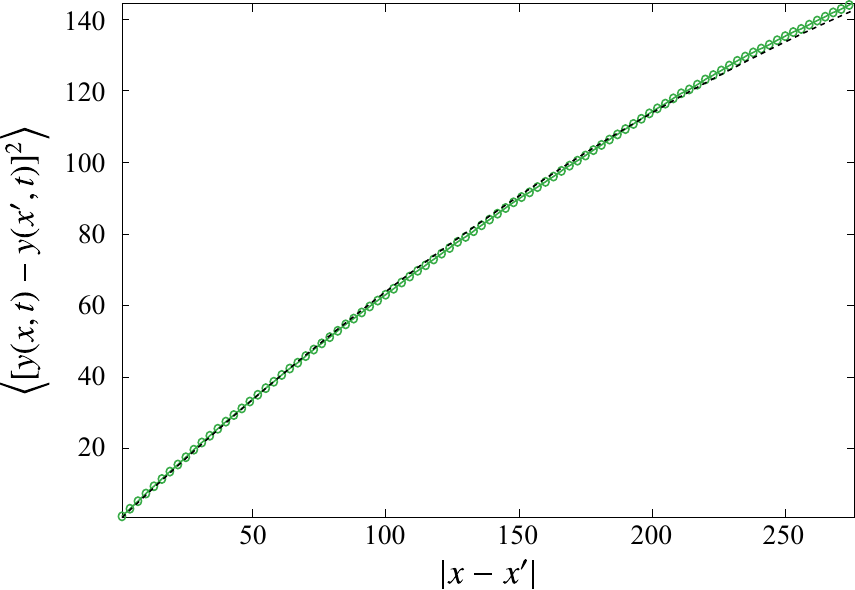}
\end{center}
\caption{Mean-square transverse displacement of a string \(y(x,t)\) in equilibrium, given by the equal-time correlation \(\left\langle [y(x,t)-y(x',t)]^2\right\rangle\), as a function of displacement along the string, \(\lvert x - x'\rvert\). The solid (green) line shows Monte Carlo results using a dimer configuration containing a single string in a system of size \(L = 1024\). (Error bars are smaller than symbols.) The dashed (black) line shows the analytical result of \refeq{EqStringWidth}, with \(D = 1/\sqrt{2}\) determined using \refeq{EqExactStringWidth}.}
\label{FigEquilibriumStringWidth}
\end{figure}

For \(t = 0\), the sum in \refeq{EqDefineV} can be evaluated exactly, using
\begin{equation}
\sum_{n=1}^{\infty}\frac{1-\cos n \theta}{n^2} = \frac{\theta (2\pi - \theta)}{4}
\punc{,}
\end{equation}
for \(0 \le \theta \le 2\pi\), to give
\begin{equation}
\label{EqStringWidth}
\left\langle [y(x,t)-y(0,t)]^2\right\rangle = V_L(x,0) = D x \left(1 - \frac{x}{L}\right)\punc{,}
\end{equation}
for \(0 \le x \le L\). Comparison with the microscopic result of \refeq{EqExactStringWidth} fixes \(D = 1/\sqrt{2}\). In \reffig{FigEquilibriumStringWidth}, both \refeq{EqStringWidth} and the value of \(D\) are confirmed using results of Monte Carlo (MC) simulations of the dimer model.

\begin{figure}
\begin{center}
\includegraphics[width=\columnwidth]{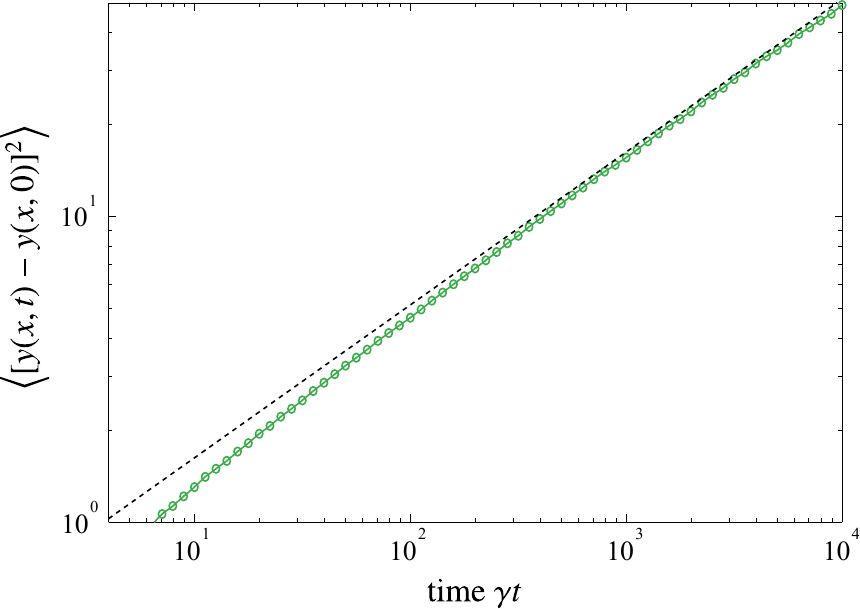}
\end{center}
\caption{Growth of mean-square transverse displacement of a string, \(\left\langle [y(x,t)-y(x,0)]^2\right\rangle\), at a fixed position \(x\), as a function of time \(t\).  The solid (green) line shows Monte Carlo results using a dimer configuration containing a single string in a system of size \(L = 1024\). (Error bars are smaller than symbols.) The dashed (black) line shows the analytical result of \refeq{EqStringDisplacementVariance2}, using the values of \(D\) and \(\Lambda\) fixed using \refeqand{EqExactStringWidth}{EqFlippability}.}
\label{FigEquilibriumStringGrowth}
\end{figure}
In calculating \(V_L(0,t)\) for small \(t / L^2\), the sum can be replaced by an integral,
\begin{equation}
\sum_{n=1}^{\infty}\frac{1-\ee^{-\frac{1}{2}n^2 \tau}}{n^2} \simeq \int_0^\infty \dd n\, \frac{1-\ee^{-\frac{1}{2}n^2 \tau}}{n^2} = \sqrt{\frac{\pi}{2}\tau}\punc{,}
\end{equation}
giving
\begin{equation}
\label{EqStringDisplacementVariance2}
\left\langle [y(x,t)-y(x,0)]^2\right\rangle = V_L(0,t) = D \sqrt{\frac{2\Lambda t}{\pi}}\punc{,}
\end{equation}
for \(\Lambda t \ll L^2\). As argued in the main text, \refeq{EqFlippability} can be used to fix \(\Lambda = 2(\!\sqrt{2} - 1)\gamma\). This result, including the value of \(\Lambda\), is confirmed using MC results in \reffig{FigEquilibriumStringGrowth}.

At large \(t\) and \(x = 0\), the first term in \refeq{EqDefineV} dominates, and so
\begin{equation}
\left\langle [y(x,t)-y(x,0)]^2\right\rangle = V_L(0,t) = \frac{D}{L}\Lambda t\punc{,}
\end{equation}
for \(\Lambda t \gg L^2\).

\section{Limiting forms of dynamical correlations}
\label{AppDynamicalCorrelations}

In the thermodynamic limit, \(L\rightarrow\infty\), \refeq{EqDefineV} can be replaced by
\beq{EqDefineV2}
V_\infty(x,t) = \frac{D}{\pi}\int_{-\infty}^\infty \dd k \, \frac{1-\ee^{\ii k x}\ee^{-\frac{1}{2}\Lambda t k^2}}{k^2}\punc{,}
\eeq
where the Cauchy principal value is to be taken, and the correlation functions can be expressed in terms of
\begin{equation}
\label{EqFullCorrelationFunction}
C\sub{s}(\qv,t) = \int_{-\infty}^{\infty} \dd x\, \ee^{-\ii q_x x} \ee^{-\frac{1}{2}q_y^2 V_\infty(x,t)}\punc{.}
\end{equation}

For \(t = 0\), one has \(V_\infty(x,0) = D \lvert x\rvert\), and so
\begin{equation}
\label{EqCs0}
C\sub{s}(\qv,0) = \frac{D q_y^2}{q_x^2 + \frac{1}{4}D^2 q_y^4} = \frac{Dq_y^2}{\omega\sub{s}(\qv)}\punc{,}
\end{equation}
where \(\omega\sub{s}(\qv) = q_x^2 + \kappa_y^4\), with \(\kappa_y = \sqrt{\frac{D}{2}}q_y\).

\begin{widetext}
For small but nonzero \(t\), consider the difference
\begin{align}
C\sub{s}(\qv,t) - C\sub{s}(\qv,0) &= \int_{-\infty}^{\infty} \dd x\, \ee^{-\ii q_x x}
\left[ \ee^{-\frac{1}{2}q_y^2 V_\infty(x,t)} - \ee^{-\frac{1}{2}q_y^2 V_\infty(x,0)}\right]\\
&\simeq -\frac{1}{2}q_y^2\int_{-\infty}^{\infty} \dd x\, \ee^{-\ii q_x x}\ee^{-\frac{1}{2}q_y^2 V_\infty(x,0)}
\left[ V_\infty(x,t) - V_\infty(x,0)\right]\\
&= -\frac{D q_y^2}{2\pi}\int_{-\infty}^{\infty} \dd x \, \ee^{-\ii q_x x}\ee^{-\frac{1}{2}q_y^2 D \lvert x \rvert} \int_{-\infty}^\infty \dd k \, \ee^{\ii k x} \frac{1 - \ee^{-\frac{1}{2}\Lambda t k^2}}{k^2}
\punc{.}
\end{align}
For \(\Lambda t \ll q_y^{-4}\), the integral over \(k\) can be replaced by \(\frac{1}{2}\Lambda t \times 2\pi\delta(x)\), and so the result is
\begin{equation}
C\sub{s}(\qv,t) - C\sub{s}(\qv,0) \simeq -\frac{1}{2}D q_y^2 \Lambda t\punc{.}
\end{equation}
Using \refeq{EqCs0} gives
\begin{equation}
\label{EqStringDynamicCorrns1}
\frac{C\sub{s}(\qv, t)}{C\sub{s}(\qv, 0)} \simeq \exp \left[-\frac{1}{2}\Lambda t \omega\sub{s}(\qv)\right]\quad\text{for \(\Lambda t \ll q_y^{-4},L^2\).}
\end{equation}

For large time (but with \(\Lambda t \ll L^2\)), one can use the saddle-point approximation. The closest saddle points to the real line are at \(x = \pm \ii x_0\), where
\begin{equation}
x_0 = \sqrt{2 \Lambda t} \operatorname{erfi}^{-1}\left(\frac{q_x}{\kappa_y^2}\right)\punc{,}
\end{equation}
and the resulting correlation function is
\begin{equation}
\label{EqStringDynamicCorrns2}
\frac{C\sub{s}(\qv, t)}{C\sub{s}(\qv, 0)} \simeq \left(\frac{\pi}{2}\right)^{3/4}\frac{(\Lambda t)^{1/4}\omega\sub{s}(\qv)}{\kappa_y^3\Xi_0^{1/2}}
\exp\left(-\Xi_0\kappa_y^2\sqrt{\frac{2}{\pi}\Lambda t}\right)\quad\text{for \(q_y^{-4} \ll \Lambda t \ll L^2\),}
\end{equation}
where $\Xi_0 = \exp [ (\operatorname{erfi}^{-1}\frac{q_x}{\kappa_y^2})^2]$. (The precise condition for the validity of the saddle-point approximation also involves \(q_x\), but the quoted inequality is always sufficient, and also necessary except for very large \(q_x / \kappa_y^2\).)

\end{widetext}

\section{Numerical results for dynamical correlations}
\label{AppNumerics}

Our analytical results for the correlations are based on a coarse-grained description of the strings, and so are expected to be quantitatively accurate only for small \(\qv\). According to \refeq{EqStringDynamicCorrns2}, however, the stretched-exponential decay is visible only for \(\Lambda t \gg q_y^{-4}\), a time scale that grows rapidly as \(q_y\) is decreased. The value \(\qv = \{\frac{\pi}{16},\frac{\pi}{4}\}\) used in Fig.~2 of the main text is chosen to show the stretching most clearly on time scales accessible in the continuous-time MC simulations. (Because we have neglected the periodicity in the \(y\) direction, we also require \(L q_y \gg 1\).)

\begin{figure*}
\begin{center}
\includegraphics[width=\textwidth]{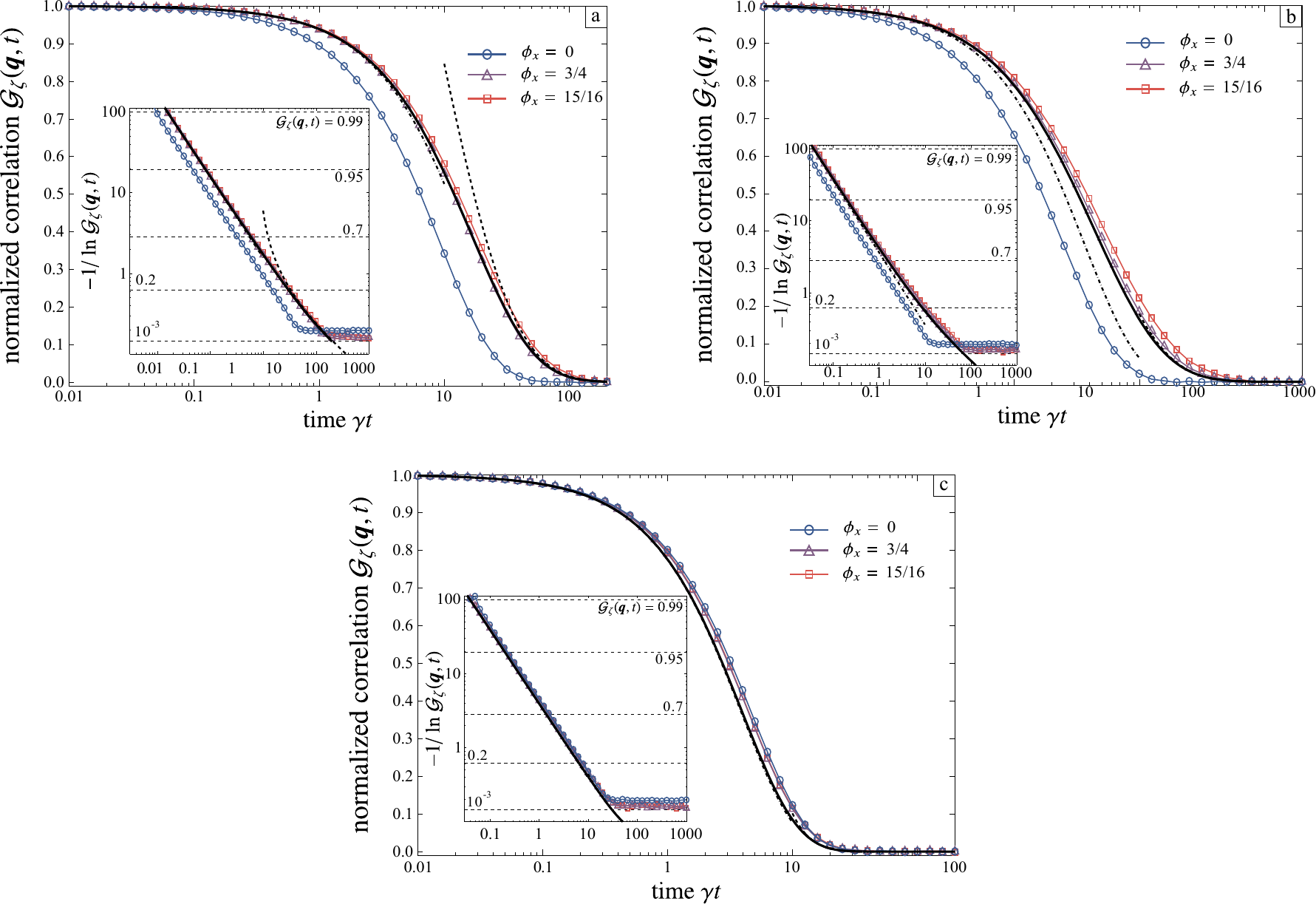}
\end{center}
\caption{Normalized height correlations \(\scG_\zeta(\qv,t) = G_\zeta(\qv,t)/G_\zeta(\qv,0)\) at flux \(\phiv = \{\phi_x,0\}\) and wavevectors (a) \(\qv = \{\frac{\pi}{8},\frac{\pi}{8}\}\), (b) \(\qv = \{\frac{\pi}{4},\frac{\pi}{4}\}\), and (c) \(\qv = \{\frac{\pi}{4},\frac{\pi}{16}\}\). In each case, the symbols show MC results with system size \(L = 256\) (error bars are smaller than symbols), while the thick (black) line shows the theoretical prediction for the large-flux limit (i.e., close to the maximum \(\phi_x = 1\)). The short- and long-time limits are shown with dot-dashed and dashed lines, respectively. Insets: Same data with double-logarithmic vertical scale; stretched exponentials appear as straight lines. The saturation at long times is an artifact resulting from the statistical uncertainty.}
\label{FigCorrelations2}
\end{figure*}

Results for other values of \(\qv\) are shown in \reffig{FigCorrelations2}. As expected, the quantitative accuracy of the analytical results decreases as \(\lvert\qv\rvert\) is increased. Consistent with \refeq{EqStringDynamicCorrns2}, clear evidence of stretched-exponential decay is visible, as a decreased slope on a double-logarithmic scale, only for the larger values of \(q_y\).

\section{Persistence time}
\label{SecPersistenceTime}

Given a trajectory, a plaquette is referred to as ``persistent'' if it has not flipped at any point during the trajectory. At each time \(t\) during the trajectory, the persistence is defined as the proportion of plaquettes that are persistent, i.e.,
\begin{equation}
p(t) = \frac{1}{N}\sum_{\rv}
\begin{cases}
1 & \text{if plaquette \(\rv\) is persistent}\\
0 & \text{otherwise.}
\end{cases}
\end{equation}
The persistence time \(\tau\sub{p}\) is the average, over starting configurations and trajectories, of the integral of the persistence,
\begin{equation}
\tau\sub{p} = \left\langle \int_0^\infty\!\! \dd t\, p(t) \right\rangle\punc{.}
\end{equation}

According to \refeq{EqStringDisplacementVariance}, in the thermodynamic limit the typical spread of \(y(x,t)\) is proportional to \(t^{1/4}\). The typical time to reach a plaquette at a distance \(\ell\) from the starting position of the string is therefore \(\ell^4\). Since a plaquette can only flip when a string is nearby, its persistence time is given by the time at which a string first reaches it. The linear density of strings is \(\sim \theta\), and so the typical distance from a plaquette to the nearest string is \(\sim\theta^{-1}\). The typical persistence time \(\tau\sub{p}\) is therefore \(\sim\theta^{-4}\).

\begin{figure*}
\begin{center}
\includegraphics[width=0.65\textwidth]{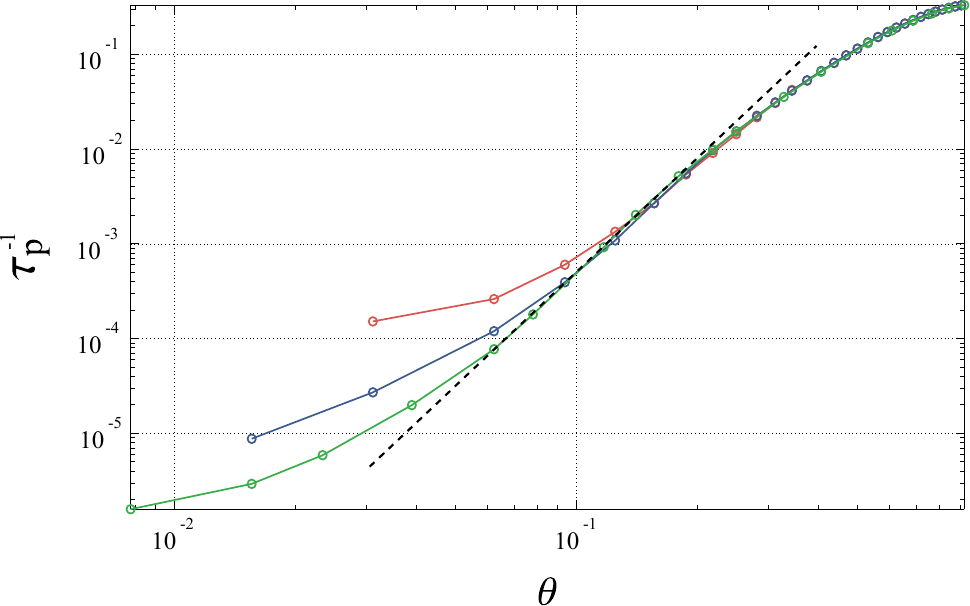}
\end{center}
\caption{Inverse persistence time \(\tau\sub{p}^{-1}\) as a function of deviation from maximum flux, \(\theta = 1 - \phi_x\). The symbols show Monte Carlo results for system sizes \(L=64\) (red), \(128\) (blue), and \(256\) (green). (Error bars are smaller than symbols.) The dashed (black) line shows the analytical prediction \(\tau\sub{p} \sim \theta^{-4}\), which applies for \(0.05 \lesssim \theta \lesssim 0.2\) (close, but not too close, to maximum flux). For larger \(\theta\), the density of strings is sufficiently high that their interactions become important, and the picture of independent strings breaks down. When \(\theta \lesssim L^{-1}\), finite-size effects become important.}
\label{FigPersistenceIntegral}
\end{figure*}

In \reffig{FigPersistenceIntegral}, this prediction is confirmed using MC simulations for intermediate values of \(\theta\). For larger \(\theta\), the string density is sufficiently high that interactions between strings become important, while for the smallest values of \(\theta\), the distance from a plaquette to its nearest string is bounded by the system size \(L\).

\end{document}